# 感染症流行時におけるスマートフォンを用いた大学生の身体活動量分析

西山 勇毅[4,a)] 柿野 優衣[1] 中 縁嗣[3] 野田 悠加[2] 羽柴 彩月[1] 山田 佑亮[1]
佐々木 航[1] 大越 匡[1] 中澤 仁[2] 森 将輝[2] 水鳥 寿思[3] 塩田 琴美[3]
永野 智久[5] 東海林 祐子[1] 加藤 貴昭[2]

**概要**：新型コロナウイルス感染症（COVID-19）の世界的な感染拡大に伴い，多くの大学ではキャンパス内での感染予防のために，キャンパスの封鎖とインターネット越しに授業を配信するオンライン授業が導入され，学生達は自宅から授業に参加している．オンライン中心の新しい生活様式は，感染リスクを低減できる一方で，運動不足による二次的な健康被害を引き起こす可能性が危惧されている．学生の健康状態の把握は大学にとって重要であるが，新しい生活様式における大学生の身体活動量の実態は明らかになっていない．そこで本研究では，日常生活中の身体活動データ（歩数と6種類の行動種別）を大学生が所有するスマートフォンを用いて自動収集し，大学生の身体活動量の変化を明らかにする．身体活動データは，大学の必修授業（体育）を履修する大学1年生305名から10週間収集した．その結果，COVID-19流行下での平均歩数(3522.5歩)は流行前の平均歩数(6474.87歩)より45.6%低下し，特に平日の歩数はCOVID-19流行前と比べる51.9%低下していることが明らかになった．また時刻別では，通学・授業中の時間帯において歩数の低下と静止時間の増加が見られ，通学や教室移動に伴う日常生活中の無意識の運動機会の減少が，平日の身体活動量の低下を招いている可能性が示唆された．

**キーワード**：モバイルセンシング，COVID-19，身体活動量分析，歩数，行動認識

# Physical Activity Analysis of College Students During the COVID-19 Pandemic Using Smartphones

Yuuki Nishiyama[4,a)] Yuui Kakino[1] Enishi Naka[3] Yuka Noda[2] Satsuki Hashiba[1] Yusuke Yamada[1] Wataru Sasaki[1] Tadashi Okoshi[1] Jin Nakazawa[2] Masaki Mori[2] Hisashi Mizutori[3] Kotomi Shiota[3] Tomohisa Nagano[5] Yuko Tokairin[1] Takaaki Kato[2]

*Abstract*: Owing to the pandemic caused by the coronavirus disease of 2019 (COVID-19), several universities have closed their campuses for preventing the spread of infection. Consequently, the university classes are being held over the Internet, and students attend these classes from their homes. While the COVID-19 pandemic is expected to be prolonged, the online-centric lifestyle has raised concerns about secondary health issues caused by reduced physical activity (PA). However, the actual status of PA among university students has not yet been examined in Japan. Hence, in this study, we collected daily PA data (including the data corresponding to the number of steps taken and the data associated with six types of activities) by employing off-the-shelf smartphones and thereby analyzed the changes in the PA of university students. The PA data were collected over a period of ten weeks from 305 first-year university students who were attending a mandatory class of physical education at the university. The obtained results indicate that compared to the average number of steps taken before the COVID-19 pandemic (6474.87 steps), the average number of steps taken after the COVID-19 pandemic (3522.5 steps) has decreased by 45.6%. Furthermore, the decrease in commuting time (7 AM to 10 AM), classroom time, and extracurricular activity time (11 AM to 12 AM) has led to a decrease in PA on weekdays owing to reduced unplanned exercise opportunities and has caused an increase in the duration of being in the stationary state in the course of daily life.

## 1. はじめに

新型コロナウイルス感染症（COVID-19）が世界的に感染拡大している．COVID-19 は，新型コロナウイルス（SARS-CoV-2）に感染し，発症すると発熱や咳，息苦しさが現れ，感染が肺に及んで肺炎を引き起こす感染症であり[1]，嗅覚や心肺機能異常，継続的な倦怠感などの後遺症[2],[3],[4]も指摘されている．ジョンズ・ホプキンス大学のまとめ[5]によると，2020 年 8 月 29 日時点で世界中で 2400 万人以上が感染し，死者は 83 万人以上にのぼり，日本国内においても 2020 年 8 月 29 日時点で累計感染者数が 67000 人に達している．COVID-19 は主に飛沫・接触感染によって感染が広がるため，感染予防策として密集・密閉・密室のいわゆる三密の回避や，マスクや手洗い・消毒の徹底，感染者隔離や都市封鎖，リモートワークの導入など人の移動と接触機会を低下させる感染拡大防止策が実施されている[6]．しかしながら，SARS-CoV-2 の感染力は非常に高く，感染予防策を組み合わせながら経済活動を継続するウィズコロナ環境の長期化が見込まれている．

本研究では，COVID-19 発生前の環境を「コロナ前」，流行時を「コロナ下」，流行後を「コロナ後」と定義する．コロナ下では感染症予防のために，実空間に集まることなく，個々人の自宅からオンライン上の仮想空間に集まることで協調作業を行うリモートワークが企業や教育現場などで積極的に導入が進められている[6]．特に大学は，広域から多くの学生が集まり，教室や研究室，食堂など複数人が室内で過ごす時間が長いため三密になりやすく，感染拡大リスクが高い．そのため，大学では積極的にオンライン授業が導入され，コロナ下では大学生・大学院生の 95.4%がオンラインで授業を受講している[6]．一方，急激な外出自粛やリモートワーク化による身体活動量の低下も指摘されている．例えば，フィットネストラッカーを提供している Fitbit 社は，コロナ前とコロナ下では，平均歩数が 1000 歩以上も低下していると報告している[7],[8]．身体活動量の長期的な低下は，運動不足による二次的な健康被害の危険性が高まる．

COVID-19 の感染リスクとその予防に関する研究はこれまで数多く行われてきているが，客観的な身体活動データを用いたコロナ下における人々の身体活動量の変化，特にオンライン化による影響を大きく受けた大学生への影響はまだ明らかになっていない．COVID-19 に限らず，これまでも中東呼吸器症候群（通称 MERS），重症急性呼吸器症候群（通称 SARS）や鳥インフルエンザなど，定期的に危険な感染症が蔓延しており，コロナ後においても再度危険な感染症が流行する可能性もある．感染症の流行とその対策が，大学生に与える影響を明らかにすることは，二次災害の予防に向けた重要課題である．

本研究では，スマートフォンに搭載された歩数計と行動種別認識機能を用いて，コロナ下における日常生活中の身体活動量を計測し，コロナ前とコロナ下における身体活動量の変化を明らかにする．大学 1 年生 305 名を対象に，都道府県を跨ぐ移動やイベント開催の自粛が日本政府より国民に要請されていた，2020 年 5 月から 7 月の春学期中（10 週間）における身体活動量の計測を行い，その計測結果を分析した．

本研究の貢献は以下の通りである．

- 感染症流行下において大学 1 年生（305 名）を対象に活動量計測を行ったこと
- スマートフォンより収集した身体活動データを用いて身体活動量の変化を調査したこと
- コロナ前とコロナ下を比較すると全体で 45.6%歩数が減少し，特に通学・課外活動時間の減少に伴い，平日の歩数が 51.9%減少していることを明らかにしたこと
- 運動習慣の有無がコロナ下においても身体活動量の増減に影響を与えることを示したこと

本論文では，1 章にて本論文の全体像について述べ，2 章で関連研究をまとめ既存研究の課題を示す．3 章では，本研究の目的とアプローチを説明し，4 章にて具体的な身体活動量計測の手法や被験者，期間等について述べる．5 章では計測結果を整理し，6 章にて考察を行う．最後に，本論文の結論を 7 章にまとめる．

## 2. 関連研究

### 2.1 コロナ下における意識・行動変化

内閣府が 10,128 人に行ったインターネット調査（2020 年 5 月 25 日から 6 月 5 日）[6]によると，コロナ下において全国平均で 34.6%，東京 23 区内では 55.5%の回答者がテレワークを一度以上実施したと回答している．職種別では，教育・学習支援業のテレワーク実施率が高く，全国平均で 50.7%が一度以上のテレワークを実施している．オンライン授業を受講した割合は，大学生・大学院生では，95.4%であり，ほぼ全ての大学でオンライン授業が実施されていた．また，テレワーク率の高い東京圏では半数以上が通勤・通学時間が減少したと回答しており，約 7 割が今後も減少した通勤・通学時間と保ちたいと考え，テレワー

1 慶應義塾大学大学院政策・メディア研究科
Graduate School of Media and Governance, Keio University
2 慶應義塾大学環境情報学部
Faculty of Environment and Information Studies, Keio University
3 慶應義塾大学総合政策学部
Faculty of Policy Management, Keio University
4 東京大学生産技術研究所
Institute of Industrial Science, The University of Tokyo
5 横浜商科大学商学部
Department of Commerce, Yokohama College of Commerce
a) yuukin@iis.u-tokyo.ac.jp

クの希望率も高い．COVID-19の流行を受けて，テレワーク・オンライン授業に対する人々の意識は大きく変化しており，コロナ下だけでなくコロナ後においても，テレワーク・オンライン授業はある程度継続される可能性が高い．

日本における携帯電話の普及率[9]は，2017年時点では90%以上（スマートフォンの普及率は約75%以上）であり，それらに搭載された位置情報センサや電波基地局への接続傾向から人々の行動範囲や傾向を把握する取組が広く行われ，行政の意思決定などで広く活用されている．例えば，NTTDOCOMOのモバイル空間統計[10]の結果では，コロナ下では東京23区全域の傾向として，一ヶ月前と比較して正午時点でのビジネス街での人口が減少し，逆に住宅エリアの人口が増加していることが確認されている．Yahoo! Japanのレポート[11]によると，平日の平均移動距離（同社が提供するアプリより収集した位置情報データを元に算出）は全て世代で約40%以上低下した．休日においても約20%低下したと報告している．AppleとGoogleは両社が提供するスマートフォンより収集したデータをオープンデータとして公開している[12],[13]．図1にコロナ下における移動パターンの変化を示す．緊急事態宣言が発令された，4月7日前後を境に，急激に歩行・車・公共交通の全ての行動が低下している．逆に一部地域で緊急事態宣言が解除された5月14日頃から徐々に，全ての行動が回復傾向にある．しかしながら，歩行に関しては緊急事態宣言の前の状態にまでは戻っていない．また，滞在場所の変化を図2に示す．駅や小売・娯楽施設，職場での滞在時間は減少している一方で，自宅での滞在時間は増加している．公園と日用（食品）雑貨店と薬局の滞在時間には大きな変化は見られない．フィットネストラッカーを提供するFitbit社の発表では，コロナ前とコロナ下では平均歩数が1000歩以上も低下していると報告されている[7],[8]．

このように，既存のスマートフォンやウェアラブルデバイスより収集したデータを活用することで，マクロ視点での人々の行動変化を観測できる．しかしながら，各社が提供するオープンデータは，プライバシー保護のために統計情報のみ提供されており，利用可能な属性データも限られている．その為，ユーザとその属性毎の詳細な行動変化といったミクロ視点での分析を行うことは困難である．

**2.2 日常生活中の身体活動量**

人々の生活は技術の進化や生活環境の変化に伴い，常に変化している．特に，近年の生活や仕事環境における機械化と自動化は，身体活動量の低下による健康障害にも繋がる可能性がある[14]．例えば，Van Del Ploegらの研究[15]では，World Health Organization (WHO)が定義する推奨身体活動量を満たしていたとしても，一日の合計座位時間が4時間長くなるに連れて，総死亡リスクが11%ずつ高まることを報告している．一日の覚醒時間における活動は，55から

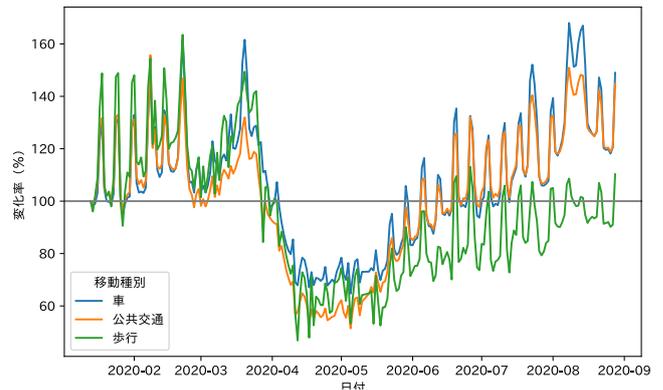

図1 コロナ下における移動パターン[12]
Fig 1. Mobility patterns under the COVID–19 pandemic [12]

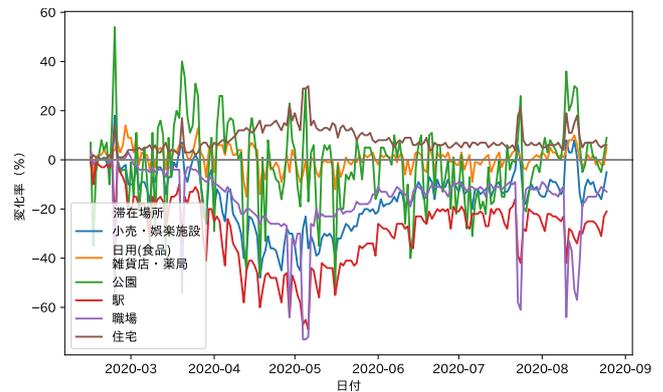

図2 コロナ下における滞在場所と滞在時間 [13]
Fig2. Visited places and duration under the COVID–19 pan- demic [13]

60%が座位行動で，35から40%が低強度身体活動（歩行など），5%がランニングなどの高強度身体活動であると報告されている[16],[17]．これらの知見から身体活動量増加の為には，座位時間を減少させ低強度身体活動量を増加させることが重要であることを示唆している．

厚生労働省がデジタル万歩計を用いた身体活動量の調査[18]によると，日本人の20〜29歳の1日の平均歩数は7308歩（男性：7904歩・女性：6711歩）と報告されている．また，国民の健康の増進の総合的な推進を図るための基本的な方針を定める「健康日本２１（第二次）」[19]では，男性は9000歩・女性は8500歩を目標に定めている．

2.1章で述べた通り，コロナ下における人々の生活は大きく変化しており，外出自粛やリモートワーク・遠隔授業の影響により，身体活動量が低下していると考えられる．特に大学では，感染拡大防止の為にキャンパスの一時封鎖や遠隔授業を導入しており，通学時間や教室移動機会，課外活動時間の減少が大学生の身体活動量にあたえる影響は大きいと考えられる．しかしながら，日本においてコロナ下における大学生の身体活動量の状況を明らかにした研究は存在しない．

**2.3 スマートフォンを用いた継続的な行動記録と分析**

モバイルクラウドセンシング（Mobile Crowd Sensing: MCS

は，学生を対象とした生活調査にも多く活用されている[20]．例えば，Wang らの研究 [21 - 23]では，データ収集用アプリである StudentLife を学生のスマートフォンにインストールし，収集データから学生の健康状態や学業成績を予測している．

Ong ら[24]は，シンガポールにおける COVID-19 対策が身体活動に与える影響を調査した．調査では, 会社員（N=1824）が所有するリストバンド型ウェアラブルデバイスから睡眠・歩数データを収集し，都市封鎖前と都市封鎖中で比較を行った．その結果, 都市封鎖中は睡眠時間に大きな変化は無いが就寝時刻が遅くなる傾向があり，さらに平均歩数が都市封鎖前（9344 歩）・中（5284 歩）を比較すると約 4000 歩低下していることが明らかになった．スペインでは，Borja ら[25]が大学生（N=20）のスマートフォンとウェアラブルデバイス，アンケート調査を用いて，都市封鎖前中の身体活動量の変化を調査した．その結果, 都市封鎖は，1 日の歩数を減少（平均 8525 歩から 2754 歩に）させ, 逆にスマートフォン利用時間と睡眠時間を増加させることを明らかにした．加えてアンケート調査より，座位時間が有意に増加し，中・高強度の運動にかける時間が低下したことが分かった．米国では Huckins ら[26]が，前述の StudentLife アプリを用いてコロナ下における大学生の身体活動量とメンタルヘルス調査を行っている．Huckins らは 217 名の大学生から，2018 年より継続的に加速度センサ, 位置情報センサ, 行動種別, スマートフォン利用頻度データを記録しており，収集データを COVID-19 流行前後の学期で比較した．COVID-19 の報道が増える毎に，座位時間が増加し，訪問場所の数が減少（位置情報データより推測）し，不安やうつ症状の増加がみられた．

コロナ下に行われた調査 [24 - 26]では，共通して都市封鎖により身体活動量の低下と座位時間の長時間化が示唆されている．しかしながら，COVID-19 の感染状況とその対策は, 国と地域によって異なるため，日本で行われた COVID-19 の感染予防策が日本の大学生に与えた影響はまだ明らかになっていない．

## 3. 感染症流行時における大学生の身体活動量

大学は，数千人以上の学生が広範囲から各種交通機関を使って通学し，三密状態で教室や研究室に長時間滞在し，かつキャンパス内で頻繁に移動が起こるため，感染リスクが非常に高い環境である．そのためほぼ全ての大学（95.4%）でオンライン授業が実施（2.1 章）されるなどの感染症対策がとられ，感染症収束までは, オンライン授業と対面授業のハイブリッド授業が継続される．また既存調査より，外出自粛要請や遠隔授業・リモートワーク化などの感染対策により，日本国内における人々の活動量が低下していることが明らかになっている[10-12],[24]．これらの感染症対策は、感染を直接的に予防できる一方で，身体活動量の低下による死亡リスクの上昇[14],[15]と言った健康被害が危惧される．諸外国の調査[25],[26]では，大学生においても身体活動量の低下傾向が見られているが，感染状況や各国・地域の政策はそれぞれ異なるため，日本国内での傾向は未知である．

日本国内の大学生の身体活動量を把握することは，長期化が予想されるオンライン中心の新しい生活様式と，二時的な健康被害予防を両立する上で必要不可欠な情報である．しかしながら，コロナ下における大学生を対象とした身体活動量の現状についてはまだ調査されていない．

本研究では, COVID-19 対策として行われた外出自粛とそれに伴う遠隔授業が，大学生に与えた影響を，客観的な身体活動データを用いて分析する．客観的な身体活動データとして，MCS を用いて，学生自身が所有するスマートフォンに搭載された歩数センサと行動識別センサのデータを継続的に収集し，身体活動量の変化を分析する．具体的には，以下の 2 点について分析を行い議論する．
- コロナ前とコロナ下における身体活動量の変化
- 曜日・時刻ユーザ属性毎の身体活動量と時間の変化

## 4. 活動量計測

本稿で利用するデータセットの詳細について，データ収集の対象者と身体活動量計測手法，計測期間について述べる．

### 4.1 SFC GO：身体活動データ収集アプリケーション

慶應義塾大学湘南藤沢キャンパス（Shonan Fujisawa Campus: 以下 SFC）では，学部一年生向けの必修授業「体育 1」において，受講者全員がスマートフォンアプリ「SFC GO（SFC Going-well Online）」をインストールし，アプリを併用した授業が実施されている．本アプリは UI 開発プラットフォームの Flutter を用いて開発され，iOS と Android の両 OS（合計 99%以上のマーケットシェア率）をサポートしており，2020 年春学期より「体育 1」授業内での「コミュニケーション支援」と「客観的な運動量の計測支援ツール」，そして「日常生活中の身体活動記録ツール」として利用されている [27]．本研究では，SFC GO を用いて自動収集された身体活動データに着目し，感染症流行時における大学生の身体活動量の変化を分析する．

### 4.2 身体活動データ

モバイルセンシングフレームワークの一つである AWARE Framework (i) は，歩数や加速度, スマートフォンの利用頻度などのスマートフォンで利用可能なハード・ソフ

---
i https://awareframework.com/

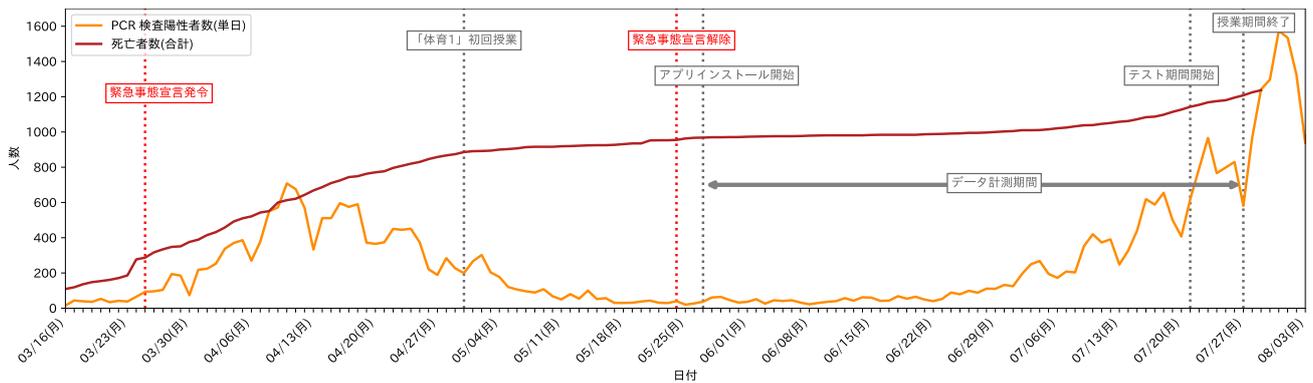

図 3　PCR 検査陽性者数と政府政策・大学授業スケジュール
Fig. 3 Positive PCR test cases, government's press releases, and the PE class schedule

トウェアセンサをユーザの入力操作無しに自動収集できる．また，前述の AWARE Framework のセンシング機能はライブラリとして，容易に既存アプリケーションに追加できる．

SFC GO では，ユーザの身体活動データとして，一分毎の「歩数」と「行動種別（歩行とランニング，自転車，自動車，静止，不明）」を AWARE Framework [28] を用いて収集している．身体活動データは，ユーザの入力操作無しに自動収集され，アプリを開閉したタイミングでサーバにアップロードされる．AWARE Framework では，上述の身体活動データを iOS と Android の両 OS が提供する API（Application Programming Interface）から取得している．しかしながら，iOS と Android では異なる身体活動の算出アルゴリズムが動いているため，本研究では iOS 端末より提供されたデータのみを分析対象とする．

AWARE Framework の iOS 版（以降 AWARE-iOS）[28] では，身体活動データを iOS 端末に標準搭載されている CoreMotion から収集している．SFC GO では，バッテリ消費を考慮し，アプリを開いたタイミングで過去の身体活動データを遡って収集した．CoreMotion は一週間分の身体活動データを自動的に保存しており，バックグランドでの常時計測を行わなくても身体活動データを収集できる．また識別された行動種別の信頼度合いは *low*・*medium*・*high* の三段階で取得できる(ii)．本研究では *high* と *medium* のデータを利用する．

### 4.3　被験者

本研究では，1 年生向けの体育授業「体育 1」の受講者の身体活動データを収集した．体育 1 は 1 年生の必修科目であり，クラス毎に週 1 回の授業が行われている．

当該学期における SFC GO を導入した学生は合計で 839 名であった．また本研究は，人を対象とする研究上の実験・調査における生命倫理，プライバシー保護，人権保護等の倫理審査について，慶應義塾大学 SFC 実験・調査倫理委員会から承認を受けた．アプリ導入者のうち，研究でのデータ利用に関する同意が得られた 771 名分を本研究の被験者とした．

### 4.4　期間

活動量の計測は「体育 1」の授業日程に合わせて行われた．図 3 に，「体育 1」の授業スケジュールおよび計測期間と日本国内における COVID-19 に対する PCR 検査の陽性者数を示す．体育 1 は 4 月 30 日に初回授業が開始され，その後，授業 5 週目の 5 月 28 日にアプリ配布を開始した．配布アプリを用いて，授業終了までの 10 週間分の身体活動データを収集した．

緊急事態宣言は，第一波の感染拡大が収束に向かったのを受け 2020 年 5 月 25 日に緊急事態宣言は解除されたが，対象の 10 週間は都道府県を跨ぐ移動やイベントの開催自粛要請などの対策が行われていた．また，PCR 検査陽性者数は，アプリ配布を開始した 5 月 28 日からデータ収集終了日にかけて徐々に増加していた．

### 4.5　コロナ前の身体活動データ

コロナ前の身体活動データとして，著者らが 2019 年 11 月から 12 月にかけて収集した身体活動データ[29]を利用する．コロナ前のデータセットは，本研究と同じく AWARE-iOS [28] を用いて客観的な身体活動データ（歩数や位置情報，加速度データなど）を収集している．またコロナ前のデータセットには，大学生と社会人のデータが含まれるが，本研究ではコロナ前のデータセットの中から，大学生 37 名（男性：27 名，女性：10 名）のデータを抜き出して利用する．

表 1　コロナ前・下の身体活動データセット
Table 1　Data sets of PA before/under the COVID-19 pandemic

|  | コロナ前 | コロナ下 |
| --- | --- | --- |
| 対象 | 大学生 | 大学生 |
| 人数 | 37 名 | 305 名 |
|  | 男性：27 名 | 男性 175 名 |
|  | 女性：10 名 | 女性 130 名 |
| 期間 | 2019 年 11-12 月 | 2020 年 5-8 月 |
| データ | 歩数 | 歩数・行動種別 |

---

ii https://developer.apple.com/documentation/coremotion/cmmotionactivityconfidence

## 5. 結果

SFC GO を用いて収集した身体活動データの結果について述べる．まず収集データの前処理について説明し，歩数・行動認識データの日・曜日毎の変化とコロナ前の身体活動データとの比較を行う．

### 5.1 収集データの前処理

本研究では，歩数の検知と行動認識をスマートフォンに標準搭載された機能を用いて検知・収集している．歩数記録と行動認識の感度は，OS や搭載されているセンサによって異なるため，本研究では iOS 端末のみを分析対象とした．アプリの設定不良のためにデータ収集が実行されなかった iOS 端末を除くと，合計 486 名分のデータが収集された．

歩数と行動認識データはアプリを通して，1 分毎に自動収集されており，理想的には一人あたり 1 日 1440 レコードのデータを収集できる．しかしアプリの運用は各クラスの担当教員に任されておりユーザによってデータの収集率は異なる．またアプリの強制終了やアンインストール，スマートフォンの電池切れなど，様々な理由により予定した全てのデータを計測できない可能性がある．そのため，本論文ではユーザ毎に 1 日 1440 レコードある日付を有効データとし，有効データが 7 日以上ある 305 ユーザ（男性:175 名・女性:130 名）を分析対象とした．そのうち大学公式の大学運動部活動（野球部やラグビー部，陸上部など）に所属しているユーザは 52 名，所属していないユーザは 253 名であった．

図 4 に有効なデータを提供したデバイス数を示す．「体育1」は 26 クラスから構成されており，それぞれ各曜日に分散して実施されている．そのため，アプリの有効データを提供するデバイス数は，初めの 7 日間で徐々に増加している．その後，終了日に向けて，徐々に低下している．

### 5.2 歩数の変化

図 5 にアプリ利用期間中の日付毎の平均歩数を示す．10 週間を通してユーザ平均歩数は，3522.49 歩（中央値：2201 歩，標準偏差：4031.73）であった．厚生労働省が行った調査[18]によると，日本人（20〜29 歳）の一日の平均歩数は 7308 歩と報告されている．厚生労働省の調査（7308 歩）と比較すると，本データ収集期間における被験者の平均歩数は 51.8%少ない．

本節では，属性毎，時刻毎，コロナ前・下における歩数変化についてまとめる．

#### 5.2.1 属性毎の歩数

曜日毎の平均歩数を表 2 に示す．平日の歩数は，月曜日が最も少なく（2849.32 歩），金曜日が最も多く（3380.19 歩）なった．また休日の平均歩数は，土（4828.5 歩）・日曜日（4760.34 歩）共に 4500 歩以上と平日の平均歩数（3018.99 歩）を大きく上回っている．平日（平均 3018.99 歩）と週末

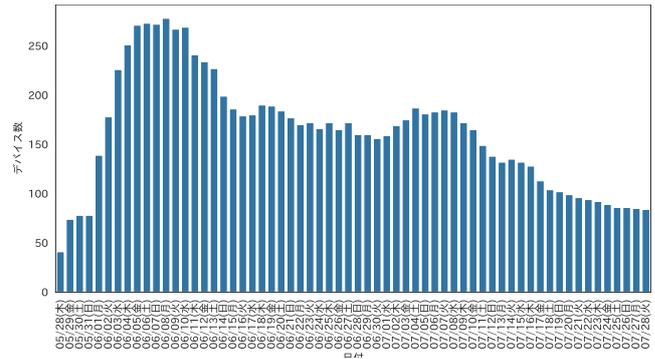

図 4 有効なデータを提供したデバイス数
Fig. 4 Number of devices that provides sufficient data

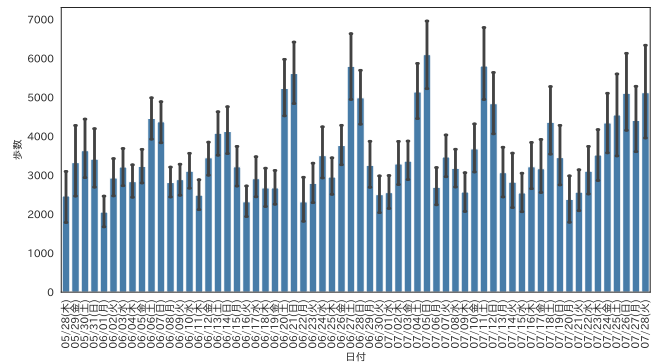

図 5 日毎の平均歩数
Fig. 5 Mean value of steps by date

表 2 曜日毎の歩数
Table 2 Mean number of steps by day of week

|   | 平均 | 標準偏差 | 中央値 |
|---|---|---|---|
| 月 | 2849.32 | 3498.93 | 1425 |
| 火 | 2943.73 | 3536.27 | 1684 |
| 水 | 3070.98 | 3518.93 | 1901 |
| 木 | 2881.39 | 3317.16 | 1601 |
| 金 | 3380.19 | 3574.04 | 2258 |
| 土 | 4828.5 | 4953.57 | 3477 |
| 日 | 4760.35 | 4894.24 | 3389 |

（平均 4797.39 歩）の歩数に対して Welch の t 検定を行った（図 6(a)）．その結果，$p<0.01$ となり，二群間に有意差が認められた．

図 6(b)に性別毎の歩数データ（男性 175 名，女性 130 名）を示す．男性の平均歩数は 3650.62 歩（SD：3963.49 歩，中央値：2439 歩）と，女性は 3348.18 歩（SD：4116.83 歩，中央値：1829 歩）であり，平均歩数は男性の方が女性よりも 302.44 歩多くなった．Welch の t 検定の結果，二群間に有意差($p<0.01$)が認められた．

図 6(c)に，大学運動部活動に所属している学生（大学運動部活動所属者）とそれ以外の学生（大学運動部活動非所属者）の日毎の合計歩数の比較を示す．所属あり（52 名）の平均歩数は 4903.13（SD:4257.89，中央値：4074）歩，逆に所属

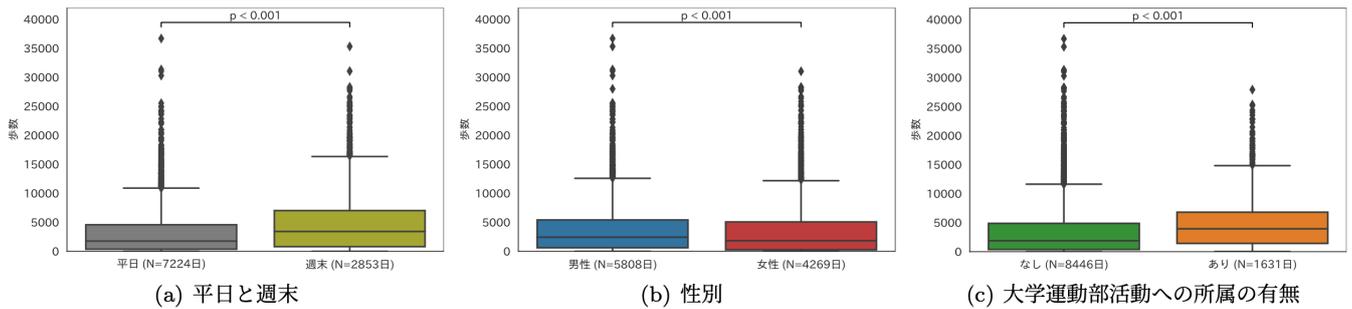

(a) 平日と週末　　(b) 性別　　(c) 大学運動部活動への所属の有無

図 6　属性毎の歩数比較
**Fig. 6**　Comparison of daily steps by attributes

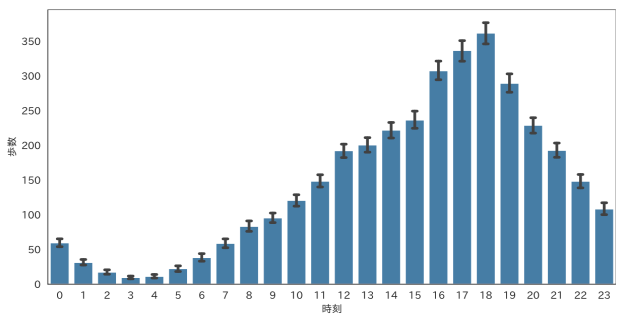

図 7　コロナ下における時刻毎の平均歩数
**Fig. 7**　Mean value of steps by the time of day under the COVID-19 pandemic

なし（253 名）は 3604.73（SD:3985.98，中央値：2319）歩であった．所属あり・なし群に対して，Welch の t 検定を行った結果，($p<0.01$)となり，二群間に有意差が示された．

#### 5.2.2 時刻毎の平均歩数

図 7 に，コロナ下における時刻毎の平均歩数を示す．平均歩数は，深夜 3 時台に最小（9.65 歩），18 時台に最大（361.78 歩）となる単峰性のデータとなった．

図 8 に，5.2.1 章にて有意差の見られた属性毎の各時刻における平均歩数を示す．平日・週末(図 8(a))と性別(図 8(b))，大学運動部活動非所属者は，全体の時刻毎の平均歩数（図 7）と同様に単峰性データであった．一方で，大学運動部活動所属者（図 8(c)）は，8 時と 12 時，18 時にピークのある多峰性データであった．

各属性の同時刻間の歩数に対して Welch の t 検定を行い，有意差を検証した．平日・週末間（図 8(a)）では，9-20 時台にかけて有意差（$p<0.01$）が得られた．また男女間（図 8(b)）では，21-2 時の深夜の時間帯にかけて有意差（$p<0.001$）が確認された．大学運動部活動への所属の有無間（図 8(c)）では，5-14 時と 16-21 時にかけて有意差（$p<0.001$）が示された．

#### 5.2.3 コロナ前・下の歩数比較

図 9(左)にコロナ前・下における一日の合計歩数の示す．コロナ前・下の一日の合計歩数の二群間に対して，t 検定を行った結果，有意差（$p<0.01$）が認められた．コロナ前・下の 1 日の平均歩数は，コロナ前の 6474.87 歩に対してコロナ下では 3522.49 歩となり，コロナ下ではコロナ前と比べて平均歩数が 45.6%減少した．

ついで，図 9（右）にコロナ前・下における時刻毎の歩数を示す．コロナ下の歩数は，5.2.1 で述べた通り単峰性のデータであるが，コロナ前のデータは 9 時と 12 時，14 時，18 時にピークが出現する多峰性のデータであった．コロナ前・下の歩数間で t 検定を時刻毎に行った結果，早朝 7 時台から深夜 0 時台の間に有意差（$p<0.01$）が得られ，コロナ下では就寝時間帯以外は歩数が少ない傾向があった．

図 10 にコロナ前・下における平日と週末の 1 日の歩数を示す．コロナ前（平日・週末），コロナ下（平日・週末）の 4 群間に対して，Tukey-Kramer の多重比較検定を行った．その結果，コロナ前（平日）とコロナ前（週末）以外の群間には，有意差（$p<0.01$）が認められた．コロナ前は平日と週末間の歩数変化は認められなかったが，コロナ下での平日の歩数は休日より有意に低下した．特に平日の平均歩数は，コロナ下（平均：3020.07 歩）ではコロナ前（平均：6278.6 歩）と比較して 51.9%（平均：3258.53 歩）減少した．

### 5.3 行動時間の変化

本節では，まず期間全体を通しての行動種別毎の行動時間をまとめ，次いでユーザ属性・時刻毎の行動時間について整理する．

図 11 に一日の行動種別毎の時間を示す．一日平均の約 1082 分（18 時間）は静止で，約 118 分（2 時間弱）歩行であった．216 分（3.6 時間）弱は，iOS の行動認識アルゴリズムが判定できなかったため不明であった．残りの約 20 分はランニング（約 16 分）・自動車（約 2 分）・自転車（約 2 分）に分類された．

#### 5.3.1 属性毎の変化

図 12 に，5.2.1 章にて有意差の見られた，3 つの属性（週末と平日・性別・大学運動部活動への所属の有無）毎の 1 日の行動時間を比較する．t 検定の結果，全属性の 5 つ全ての行動種別（静止・歩行・ランニング・自動車・自転車）の全てにおいて $p<0.01$ の有意差が見られた．

図 12(a)に示すように，平日は週末と比べて有意に静止時間が増加し，逆にその他の行動時間が短くなった．また性別間（図 12(b)）では，男性と比べて女性の静止時間が有意に

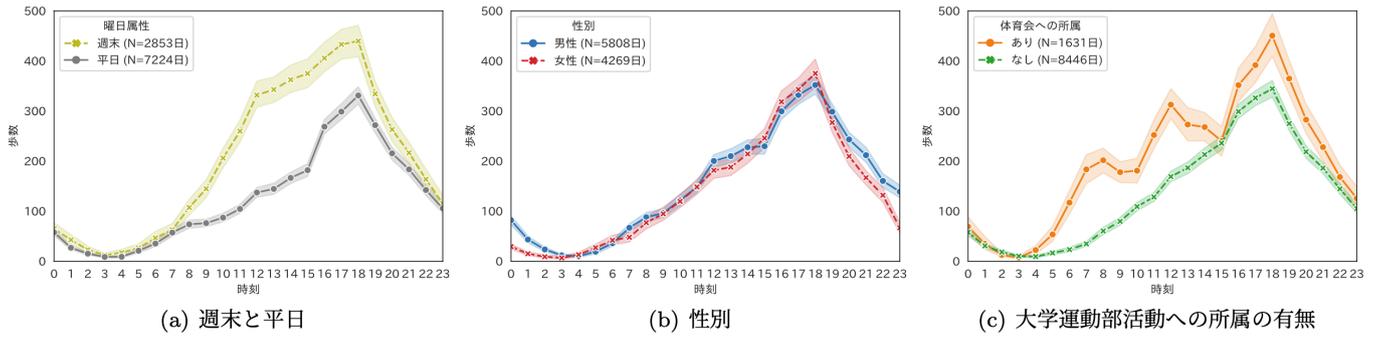

(a) 週末と平日　　(b) 性別　　(c) 大学運動部活動への所属の有無

図 8　属性毎の各時刻における平均歩数

Fig. 8　Mean value of steps by the time of day and each attribute

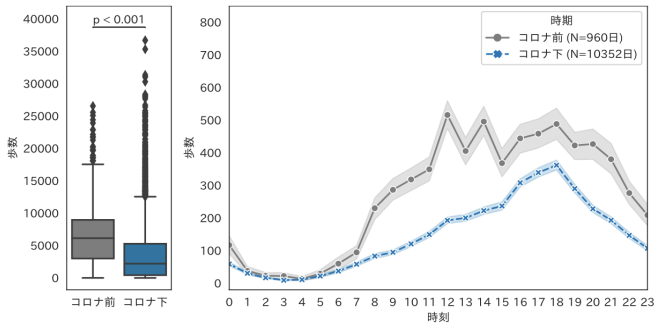

図 9　コロナ前・下における歩数

Fig. 9　Number of steps before and under the COVID-19 pandemic

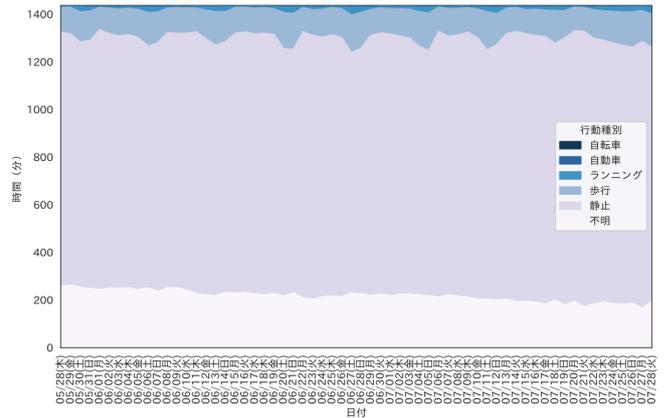

図 10　コロナ前・下における平日・週末の一日の合計歩数

Fig. 10　Number of daily steps on weekdays and weekends before and under the COVID-19 pandemic

長くなり，その他の行動時間が短くなった．大学運動部活動への所属の有無（図 12(c)）でも同様に，大学運動部活動所属者と非所属者を比べると有意に静止時間が長くなり，その他の行動時間が短くなった．

**5.3.2 時刻毎の変化**

各時刻における行動種別（6 種類）の平均時間を図 13 に示す．歩行時間は，早朝から徐々に増加し，18 時頃をピークに深夜（4 時）に向けて減少している．逆に静止時間は，深夜 4 時をピークにして，夕方に向けて徐々に減少する．ランニング・自転車・自動車の時間は静止・歩行と比べると短い時間ではあるが，歩行時間と同様の傾向であった．

図 11　一日の行動時間（分）

Fig. 11　PA types and times per day (minutes)

各属性の時刻毎における行動時間を図 14 と図 15，図 16 に示す．各時刻における t 検定の結果，特に週末と平日では，週末の歩行時間（図 14）が 10 時から 18 時まで有意（$p<0.01$）に多くなり，静止時間は少なくなった．また，中・高負荷の運動にあたるランニングと自転車の時間は，全ての時間帯において有意（$p<0.01$）に休日に多くなった．

歩行時間（図 15）は，早朝 4 から 5 時，15 時以外の時間は，男性の歩行時間が有意（$p<0.01$）に多くなった．静止時間は，9 時から 24 時において少なくなった．男性のランニング時間は，23 時から 1 時に有意に多くなった．大学運動部活動所属者（図 16）の歩行時間は早朝 5 時から 14 時，16 時から 24 時に有意（$p<0.01$）に多くなり，逆に歩行時間は短くなった．

## 6.　考察

計測結果をもとに，コロナ下における身体活動の変化について考察する．

### 6.1　コロナ下における身体活動量の低下

5.2 章で述べた通り，コロナ下において大学生の歩数はコロナ前と比較すると 45.6%減少した．特に平日の歩数が 51.9%有意に減少している．コロナ前・下における歩数を時刻毎に比較すると，深夜の時間帯（1 時から 6 時）に有意差は無いが，通学時間帯（7 時から 10 時）と昼間から夜間活

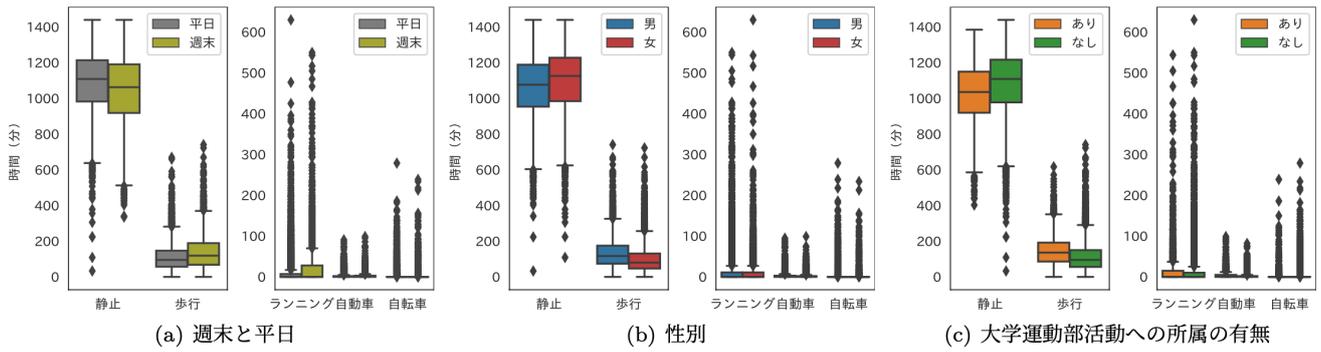

(a) 週末と平日　　(b) 性別　　(c) 大学運動部活動への所属の有無

図 12　属性毎の活動時間の比較
Fig. 12　Comparison of PA types and times by attributes

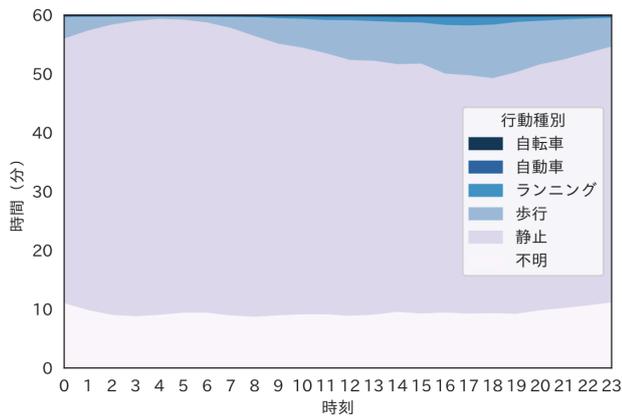

図 13　各時刻における行動種別毎の時間
Fig. 13　Amount of time for PA types at each hour of the day

動時間帯（11 時から 24 時）の歩数が大幅に低下した．通常通り大学の教室にて対面で授業を行う場合には，学生達は教室まで移動して受ける必要があるため，授業時間に合わせて起床し，徒歩や電車・バスを用いて移動する必要がある．しかしながら，計測期間中，慶應義塾大学の授業は全て遠隔講義となっており，遠隔授業では授業開始直前に起床し，移動無しに自室内から直接授業に出席できる．つまり身体活動量・時間の低下は，通学や教室移動，課外活動（サークル活動など）時に無意識確保されていた低強度の運動時間が失われたことに起因すると考えられる．一日に必要な運動量の確保には，減少した歩数を意識的に増加させる対策が必要である．

### 6.2　運動習慣の有無による身体活動量の変化

大学運動部活動所属者と非所属者と比較すると，所属者の方が有意に歩数が多くなった（図 8(c)）．これは，所属者は常日頃から各競技における個々人・チーム目標を設定し，その目標達成に向けて日頃から運動習慣や自己管理を徹底する習慣をすでに持っていることが要因であると考えられる．時刻毎の歩数パターン（図 8(c)）を見ると，コロナ前と似た通学時間と昼食，夕食時間帯に歩数のピークがある多峰性（図 8）を有する．ここからコロナ下においても，大学運動部活動所属者はコロナ下においてもコロナ前の生活リズムを大きく崩しておらず，非所属者の身体活動が低下している時間帯において，自主練習などの運動を行うことで，ある程度の身体活動量を確保したと予想される．つまり日頃の運動習慣の有無が，コロナ下においても身体活動量の低下を抑制していると考えられる．

### 6.3　身体活動促進に向けた今後の展望

全ての授業をオンラインで提供（完全オンライン授業化）すると，運動習慣がまだ備わっていない学生は無意識の身体活動時間が減少するため，結果として身体活動量が低下する．完全オンライン授業化する場合には，損失する無意識の身体活動量を確保するため，定期的に運動機会を設けるなどの工夫とそのガイドラインの作成が必要である．例えば，ラジオ体操は，夏休み中に毎朝早朝から毎日行われており，何かしらの「複数人で定時に運動を行う機会」を提供することは，生活リズムを安定させるのに効果的であると考えられる．また成人においても朝活と呼ばれる，始業前にランニングや散歩，ヨガなど早朝に趣味の時間を行う活動も流行している．

しかしながら．昨今の感染症流行状況下において，複数人が一箇所に集まって運動を実施することは感染症予防の観点から現実的に難しい．そこで，本研究で利用したアプリ（SFC GO）上で「定時（例えば 8 時）に複数人で運動（ウォーキング等）を行うイベントを仮想的に開催しリアルタイムに運動情報を共有する」など，情報技術を用いて，失った運動機会を確保することは今後の課題である．また，ユーザ属性毎に身体活動の傾向は大きく異なった．各ユーザにとって最適な身体活動の促進手法やタイミングは異なると考えられ，ユーザに応じた最適化についても今後の課題である．

### 6.4　本研究の限界点

本研究におけるデータ計測・分析手法の限界点について述べる．

#### 歩数・行動認識の精度

本研究におけるデータ計測は，スマートフォンに搭載されたセンサを用いて計測を行なった．そのため，スマートフォンを所持していない時間帯は，歩数・行動認識データ を計測できない．厚生労働省の調査 [18] では，歩数の計測に専用の歩数計を

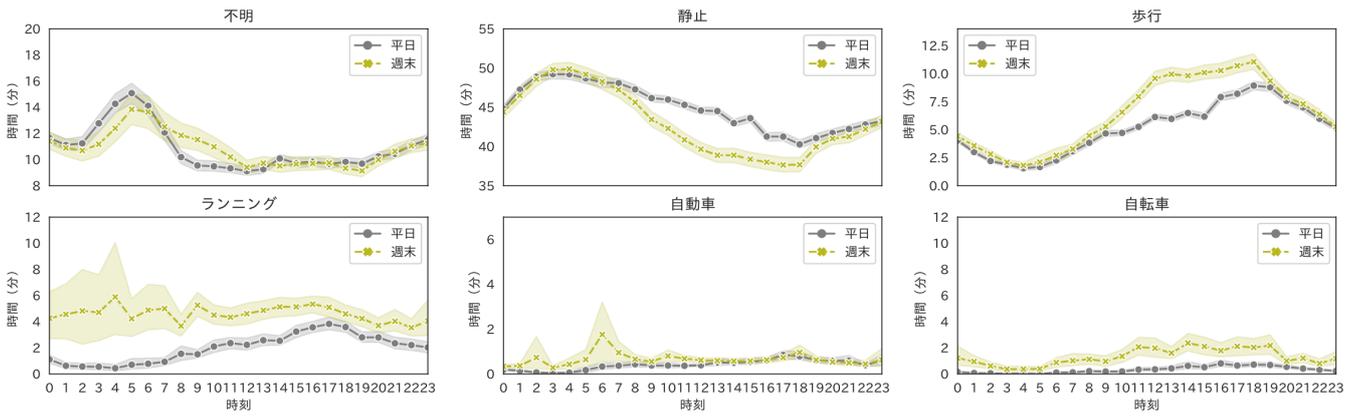

図 14 各時刻における行動種別毎の時間：平日と週末の比較
Fig. 14 Amount of time for PA types at each hour of the day: weekdays and weekends

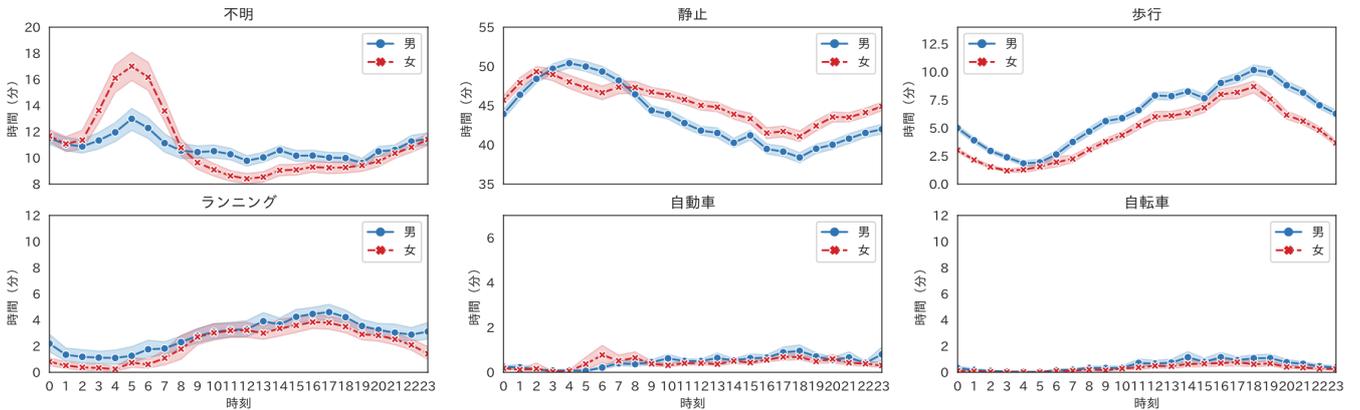

図 15 各時刻における行動種別毎の時間：性別の比較
Fig. 15 Amount of time for PA types at each hour of the day: gender

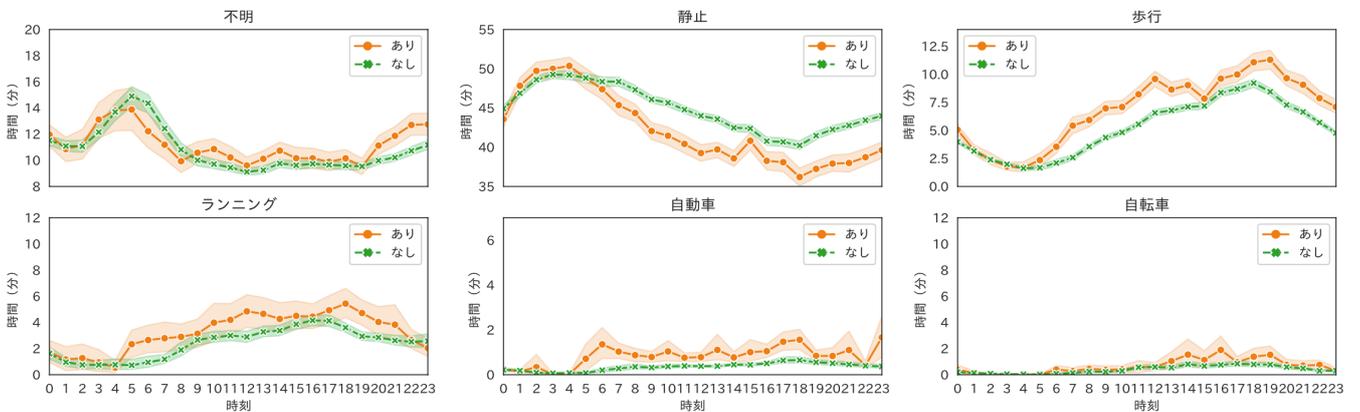

図 16 各時刻における行動種別毎の時間：大学運動部活動への所属の有無の比較
Fig. 16 Amount of time for PA types at each hour of the day: athlete and non-athlete

用いており，本研究の計測結果は，厚生労働省の調査結果よりも少ない歩数が記録されている可能性がある．

本研究では行動時間データとして，iOS が提供する行動認識 API（CoreMotion）から取得できる行動認識ラベル（静止と歩行，ランニング，自転車，自動車，不明）を活用した．本 API はスマートフォンの所有場所に限らず全ての行動を認識することができる．しかしながらスマートフォンを所持していない時間帯については，行動認識を行うことはできず「静止」に分類される．その為，本データは実態よりも静止時間が多くなる傾向がある．より正確な行動認識の実現の為には，Apple Watch 等のウェアラブルデバイスを組み合わせた計測が必要である．

また行動認識 API から出力される約 20%の行動は不明状態であり，行動種別も 6 種類と限定的であった．スマートフォンより加速度データを収集し，より高精度・多種類の行動を識別可能な行動認識アルゴリズムにかけることで，より詳細な分析が可能になると考えられる．

**季節性の排除**

本研究では，春学期の後半 10 週間の身体活動量計測を行ったため，学期末テストの影響により学期末は身体活動量が低下

している可能性がある．

気象庁の発表 (iii) によると，関東地方の梅雨入りは 6 月 11 日頃，梅雨明けは 8 月 1 日と発表されており，計測期間のほとんどが梅雨期間であった．雨天により外出を避けたことで，計測期間中の身体活動量が通常よりも少なくなった可能性もある．同時期に計測した身体活動量と比較するなど，季節性を考慮した比較は今後の課題である．

## 7. おわりに

COVID-19 の感染拡大に伴い，多くの大学では感染予防のために，オンライン授業が導入され，学生は自宅など遠隔から授業に参加している．急激なオンライン化により，人々の身体活動量は減少傾向にあり，運動不足による二次的な健康被害が懸念されている．しかしながら，コロナ下における大学生の身体活動量の実態は明らかになっておらず，コロナ下そして次の感染症流行時における大学生の健康管理のために，身体活動量の現状を明らかにする必要がある．

本研究では，感染予防策として完全オンライン授業が導入された大学において，大学一年生の身体活動データをスマートフォンから収集し，コロナ前とコロナ下の比較と，属性毎のコロナ下における身体活動量の違いを分析した．身体活動データは，大学一年生 305 人を対象に，被験者がそれぞれ所有するスマートフォンから「一分毎の歩数」と「6 種類の行動時間」を授業期間中の 10 週間，自動収集した．

その結果，コロナ前の平均歩数（6474.87 歩）とコロナ下（3522.5 歩）を比べるとコロナ下では 1 日の歩数が 45.6%低下しており，特に平日の平均歩数は，51.9%低下していることが明らかになった．コロナ下の平日と休日の活動時間を比較すると，7 時から 10 時と 11 時から 24 時における歩行時間が有意（$p<0.01$）に減少し，逆に静止時間が増加した．これは，通学や授業，課外活動における無意識の運動時間が低下したことにより平日の運動量が大幅に低下したと考えられる．また，コロナ下においては大学運動部活動所属者の歩数と中・高強度の運動時間は，非所属者と比べると有意（$p<0.01$）に増加し，歩数パターンはコロナ前の平日と同様の多峰性を示した．一般的に大学運動部活動所属者は定期的な運動機会と習慣が備わっていると考えられ，運動機会と習慣はコロナ下においても歩数の低下を軽減する効果があると推察される．

本研究では，主にコロナ下におけるモバイルクラウドセンシングを用いた大学生の身体活動量の実態調査を行った．本結果を踏まえた身体活動促進を目的とした，介入手法の開発と検証については今後の課題である．

---

iii https://www.data.jma.go.jp/fcd/yoho/baiu/sokuhou_baiu.html